\documentclass[english,aps,prl,twocolumn]{revtex4}
\usepackage{graphicx}
\usepackage{amsfonts}
\usepackage{amssymb}
\usepackage[T1]{fontenc}
\newcommand{\ket}[1]{|#1\rangle}

\begin{document}

\title{Fast Scalable State Measurement with Superconducting Qubits}

\author{Evan Jeffrey} \thanks{These authors contributed equally to this work} \affiliation{Department of Physics, University of California, Santa Barbara, California 93106-9530, USA}
\author{Daniel Sank} \thanks{These authors contributed equally to this work} \affiliation{Department of Physics, University of California, Santa Barbara, California 93106-9530, USA}
\author{J.Y. Mutus} \affiliation{Department of Physics, University of California, Santa Barbara, California 93106-9530, USA}
\author{T.C. White} \affiliation{Department of Physics, University of California, Santa Barbara, California 93106-9530, USA}
\author{J. Kelly} \affiliation{Department of Physics, University of California, Santa Barbara, California 93106-9530, USA}
\author{R. Barends} \affiliation{Department of Physics, University of California, Santa Barbara, California 93106-9530, USA}
\author{Y. Chen} \affiliation{Department of Physics, University of California, Santa Barbara, California 93106-9530, USA}
\author{Z. Chen} \affiliation{Department of Physics, University of California, Santa Barbara, California 93106-9530, USA}
\author{B. Chiaro} \affiliation{Department of Physics, University of California, Santa Barbara, California 93106-9530, USA}
\author{A. Dunsworth} \affiliation{Department of Physics, University of California, Santa Barbara, California 93106-9530, USA}
\author{A. Megrant} \affiliation{Department of Physics, University of California, Santa Barbara, California 93106-9530, USA}
\author{P.J.J. O'Malley} \affiliation{Department of Physics, University of California, Santa Barbara, California 93106-9530, USA}
\author{C. Neill} \affiliation{Department of Physics, University of California, Santa Barbara, California 93106-9530, USA}
\author{P. Roushan} \affiliation{Department of Physics, University of California, Santa Barbara, California 93106-9530, USA}
\author{A. Vainsencher} \affiliation{Department of Physics, University of California, Santa Barbara, California 93106-9530, USA}
\author{J. Wenner} \affiliation{Department of Physics, University of California, Santa Barbara, California 93106-9530, USA}
\author{A. N. Cleland} \affiliation{Department of Physics, University of California, Santa Barbara, California 93106-9530, USA}
\author{John M. Martinis} \affiliation{Department of Physics, University of California, Santa Barbara, California 93106-9530, USA}

\begin{abstract}
Progress in superconducting qubit experiments with greater numbers of qubits or advanced techniques such as feedback requires faster and more accurate state measurement. We have designed a multiplexed measurement system with a bandpass filter that allows fast measurement without increasing environmental damping of the qubits.  We use this to demonstrate simultaneous measurement of four qubits on a single superconducting integrated circuit, the fastest of which can be measured to 99.8\% accuracy in 140\,ns. This accuracy and speed is suitable for advanced multi-qubit experiments including surface code error correction.
\end{abstract}

\maketitle

With recent results showing high fidelity one and two qubit logic gates \cite{Barends:gates, Chow:2013}, superconducting qubits have become a leading candidate for experiments in large scale engineered quantum systems. Realization of complex experiments in quantum information such as error correction~\cite{Cleland:surfaceCode, Roussendorf:surfaceCode}, quantum simulation~\cite{buluta:simulation}, cluster state quantum computing~\cite{Nielsen:clusterState,Yao:ClusterState}, and measurement feedback~\cite{CampagneIbarcq:Feedback,Blok:Feedback} will require state measurements to be interleaved with coherent manipulations. For example, error correction protocols like the surface code repeatedly measure parity operators to detect and correct errors. This requires the measurement process, like the gates, to be much faster than the qubit coherence time.  In particular, the measurements must be switched on and off quickly so that the measurement channel does not continuously collapse the qubit state during coherent manipulations. Additionally, an ideal detector suitable for a large system multiplexes to many qubits without introducing correlated qubit errors.

Accurate measurement of superconducting qubits is a major challenge because the measurement apparatus introduces damping which lowers the qubit's energy relaxation time $T_1$. Transmon qubits \cite{Koch:transmon} are measured dispersively; a probe signal applied to an auxiliary linear resonator coupled to the qubit acquires a phase shift that depends on the qubit's quantum state \cite{Schuster:acStarkDephasing}. Coupling to the environment through the resonator leads to qubit damping via the Purcell effect \cite{Purcell:effect, Reed:purcell}. This places a limit on measurement speed as the resonator coupling to the environment, characterized by a leakage rate $\kappa_{r}$, must be large enough to get photons into and out of the resonators quickly, but weak enough to prevent environmental damping from lowering $T_1$. Introducing a filter between the qubit and environment eases this constraint by suppressing damping at the qubit frequency while maintaining strong coupling between the resonator and environment. Increased $T_1$ was demonstrated previously with a notch filter placed in series with the resonator, but measurement speed was not studied \cite{Reed:purcell}.

\begin{figure}
\begin{centering}
\includegraphics[width=8cm]{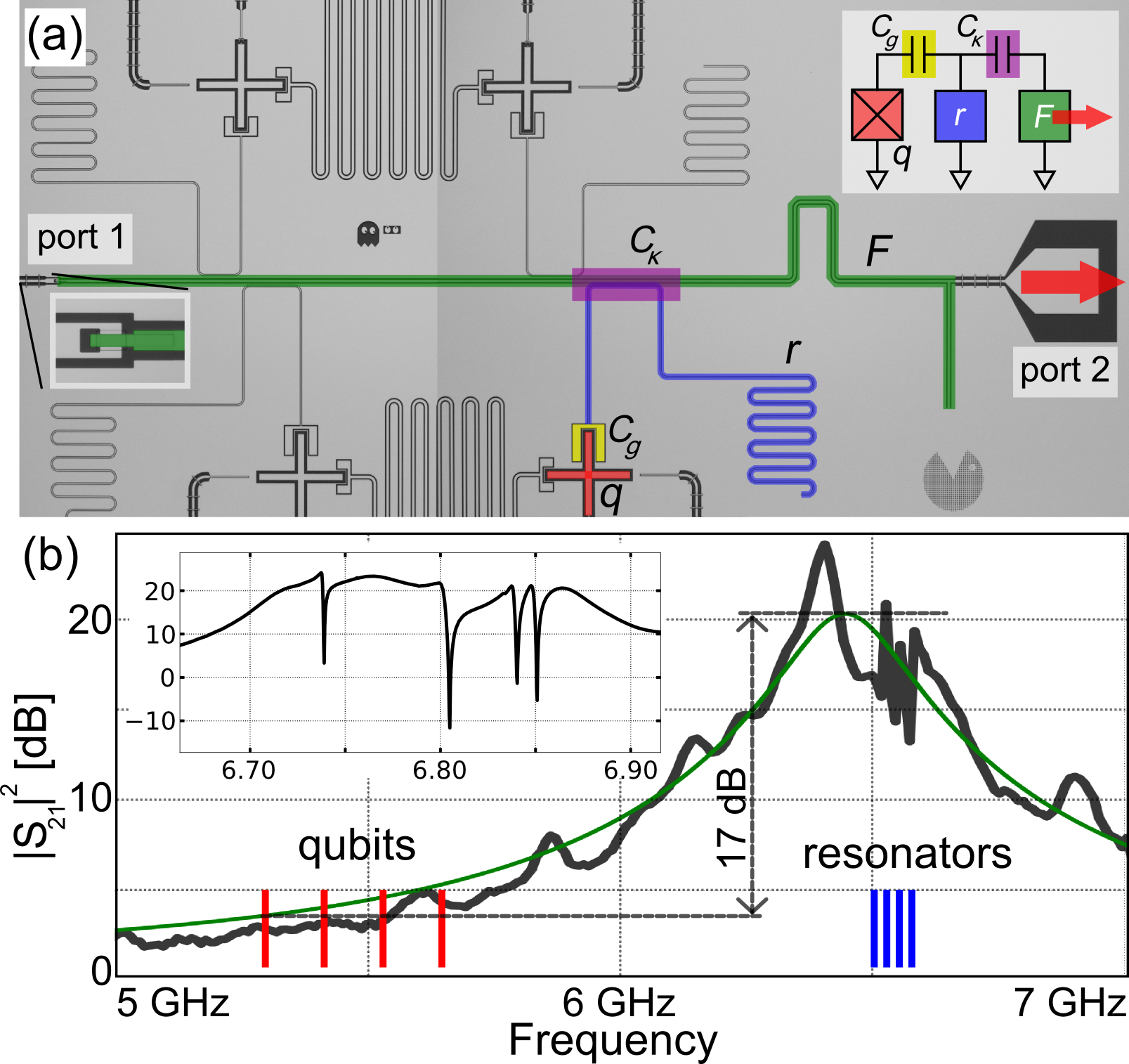}
\par\end{centering}
\caption{(color online) Device layout and frequency response. (a) Micrograph of the device with lumped element model (inset). The qubits q are coupled through a capacitor $C_g$ to the voltage anti-node of the $\lambda/4$ measurement resonator $r$. The resonators are coupled via capacitors $C_{\kappa}$ to the filter resonator $F$. The red arrow indicates how energy leaves the system. (b) Transmission spectrum $S_{21}$ of the detector. Transmission measured on a test chip is shown by the heavy (black) curve, and a Lorentzian fit is shown by the thin (green) curve. The measurement resonators are in the passband where the transmission is large, whereas the qubits are off resonance and thus protected from the environment. The inset shows a detail of the spectrum from the chip used in this experiment. Each dip in the transmission curve comes from one measurement resonator.}
\label{Fig:Figure1}
\end{figure}

Recent experiments demonstrating quantum jumps, state heralding, dressed dephasing, single quantum trajectories, and joint qubit readout have focused on a single channel of quantum information \cite{Vijay:quantumJumps, Johnson:heralding, Slichter:dressedDephasing, Murch:trajectories, Chow:jointReadout}. Furthermore these experiments used either long measurement times or qubits with coherence strongly limited by the measurement system. To make progress toward more complex experiments including a stabilized logic element or high fidelity feed-forward schemes, high measurement accuracy and speed in the transient case must be demonstrated in a multi-qubit system.

In this Letter we present a scalable qubit state detector, based on a bandpass filter, and use it to implement high speed, high accuracy multi-qubit state measurement. We introduce a design formula based on the $\kappa_rT_1$ product that characterizes the tension between the transient response rate of the measurement resonator $\kappa_r$ and the maximum qubit $T_1$ due to environmental damping. The bandpass filter design dramatically increases the $\kappa_r T_1$ limit to $\sim 6700$, with $\kappa_r=1/19\,\textrm{ns}$ in the fastest of four qubits. Based on these results, we expect that an optimized design could reach $\kappa_r = 1/10\,\mathrm{ns}$ while allowing a $T_1$ above 100\,$\mathrm{\mu}$s. We find that the bandpass filter allows four qubit simultaneous measurement with intrinsic fidelities reaching 99\% in less than 200\,ns after the start of the measurement pulse.

\begin{figure}
\includegraphics[width=8cm]{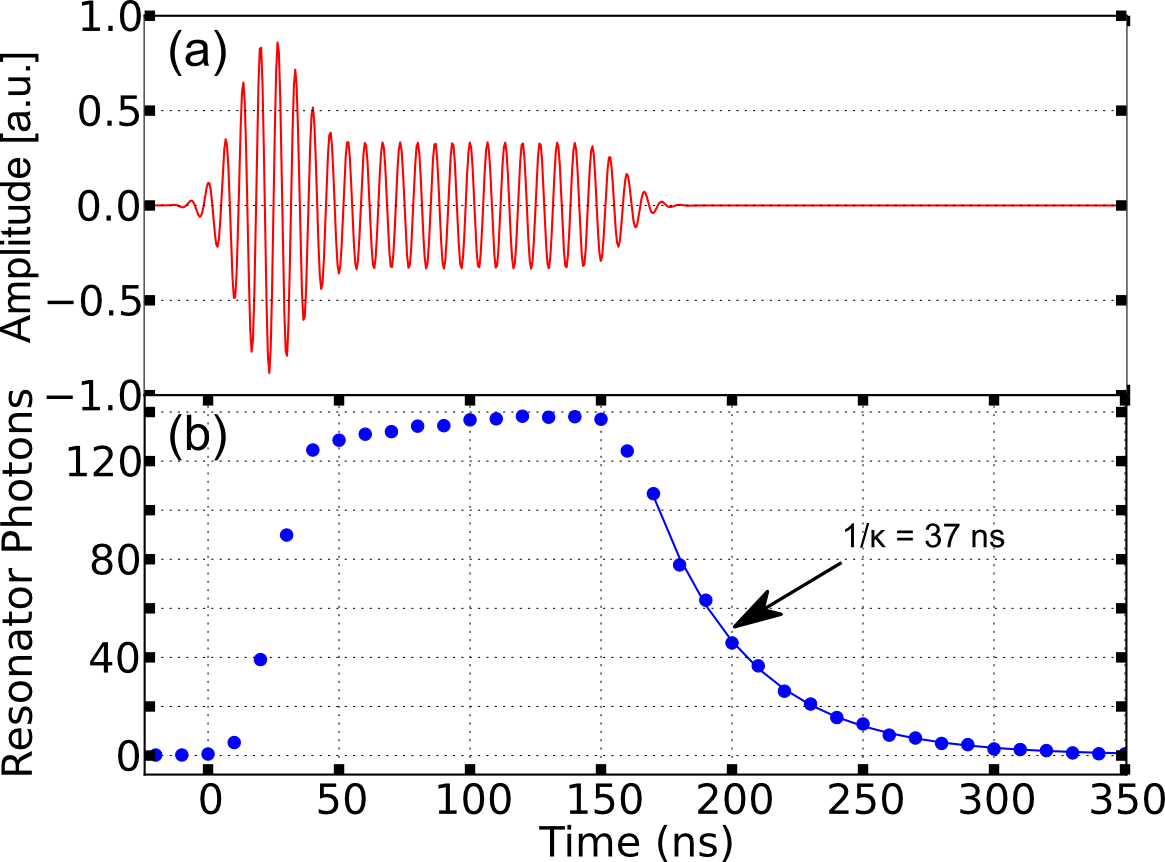}
\caption{ Measurement pulse shape and resonator photon occupation. (a) Measurement pulse produced by the AWG for a single qubit (real quadrature). (b) Time dependent population of the measurement resonator as measured by the AC Stark shift. This shows the initial 25\,ns strong drive, which quickly rings up the resonator, the sustain pulse, and the free ring-down with time constant $1/\kappa_r = 37\,\textrm{ns}$. This corresponds to a resonator $Q_r$ of 1561.}
\label{fig:readoutpulse}
\end{figure}

We achieve this fast measurement by integrating a bandpass filter into a multiplexed resonator system \cite{Chen:readout}. The device, shown in Fig.\,\ref{Fig:Figure1}(a), has four qubit/resonator pairs, designed to test the performance of different compromises between measurement speed and environmentally limited $T_1$. The filter is implemented as a quarter wave ($\lambda/4$) coplanar waveguide resonator embedded directly into the feed line. Interruption of the feed line by a capacitor (port 1 in Fig.\,\ref{Fig:Figure1}(a)) imposes a voltage anti-node, while a ground connection at a distance $\lambda/4$ imposes a voltage node. The resulting standing wave mode creates a bandpass filter as shown in Fig.\,\ref{Fig:Figure1}(b). By placing the measurement resonator frequencies but not the qubit frequencies in the pass band, the measurement resonators are strongly coupled to the environment without damping the qubits. The measurement signal couples out of the filter into the measurement environment through a tap near the voltage node. The energy leakage rate, and thus the quality factor of the filter $Q_F$, is set by the fraction of the total voltage at this tap-off point; we designed for $Q_F=30$ which gives enough bandwidth for several measurement resonators while allowing high qubit $T_1$.

\begin{figure}
\includegraphics[width=8cm]{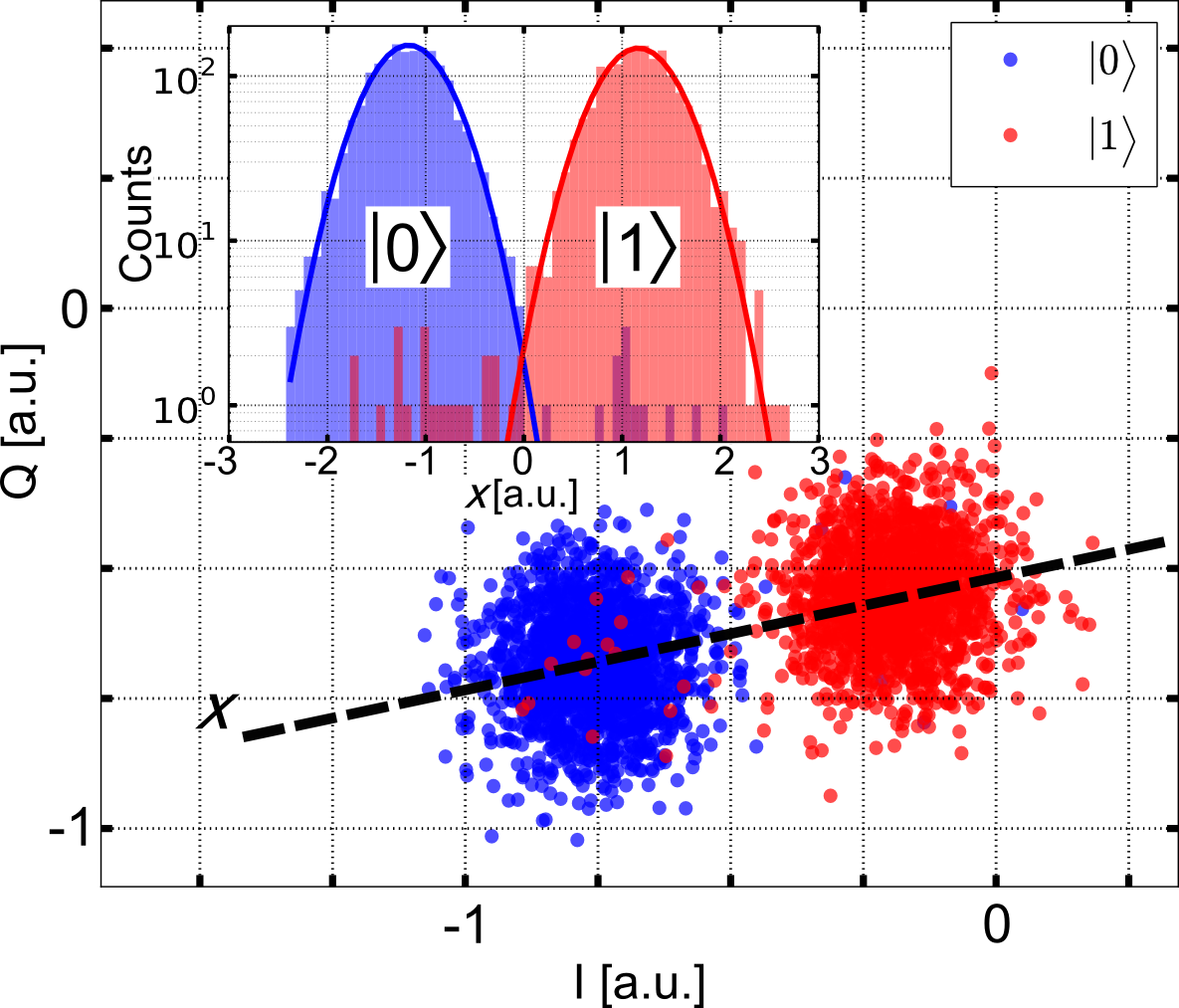}
\caption{(color online) Single shot measurement events for one qubit after 140\,ns pulse integration. Points in the wrong cluster are due to unwanted qubit transitions. The inset shows histograms of the IQ points projected onto line connecting the $\ket{0}$ and $\ket{1}$ clouds. Heavy lines are Gaussian fits to the histograms and are used for computing the separation fidelity.}
\label{Fig:Figure3}
\end{figure}

Each qubit's resonator is connected in parallel to this common filter through a capacitance $C_{\kappa}$, and each qubit is connected to its resonator by a capacitance $C_g$ to give a qubit-resonator coupling strength $g/2\pi$ between 50 and 150\,MHz.

The design was based on an analytic theory of the $\kappa_r T_1$ product, which characterizes the limit on the measurement rate $\kappa_r$ for a given environmentally limited qubit lifetime $T_1$. For the unfiltered case the product is constrained by \mbox{$\kappa_r T_1  \leq \left( \Delta/g \right) ^2$}, where \mbox{$\Delta \equiv \omega_q - \omega_r$} is the qubit-resonator detuning. The product cannot be effectively increased by raising $\Delta$ because this requires a corresponding increase in $g$ to maintain a measurable dispersive phase shift \cite{Koch:transmon}. Introducing a bandpass filter increases the $\kappa_r T_1$ product to \cite{_supplementary} \begin{equation}
\kappa_r T_1 \leq \left( \frac{\Delta}{g} \right)^2 \left( \frac{2 \Delta}{\omega_q / Q_F}\right)^2 \, . \label{eq:kappa_T1_product} \end{equation}
The second factor in Eq.\,(\ref{eq:kappa_T1_product}) allows faster measurement without lowering $T_1$; for fixed $\kappa_r$, $\Delta$, and $g$ the new limit exceeds the unfiltered one by a factor of $4Q_F^2 \Delta^2 / \omega_q^2 \approx 100$. This factor nearly matches the observed difference in system power transmission $|S_{21}|^2$ between the qubit and resonator frequencies, as shown by the vertical arrow in Fig.\,\ref{Fig:Figure1}(b). Device parameters are given in Table~\ref{Table:parameters}. With the parameters from the second row of the table and $\omega_q/2\pi = 5.5 \, \textrm{GHz}$, we compute a $T_1$ limit of $\sim 570\,\mu\textrm{s}$, which greatly exceeds the $T_1$ limit imposed by other decay channels in the experiment.

The superconducting Xmon transmon qubits were fabricated from etched Al films on a sapphire substrate as in Ref.\,\cite{Barends:XMon}. We include additional lithography and deposition steps to form Al on $\textrm{SiO}_2$ dielectric wire crossovers to suppress spurious modes on the chip and reduce parasitic inductances responsible for large unwanted frequency shifts in the filter resonance \cite{Chen:airbridges, _supplementary}.

\begin{table}
\begin{center}
\begin{tabular}{  c  c  c  c  }
\hline \hline
& $\omega_{r}/2\pi$ [GHz]	& \quad $g/2\pi$ [MHz]	& \quad $\kappa_{r}^{-1}$ [ns]	\\
\hline
$Q_1$   & \quad 6.805 (6.835)	& \quad 146 (100)	& \quad 12 (19)			\\
\hline
$Q_2$   & \quad 6.765 (6.789)	& \quad 102 (86)	& \quad 23 (37) 		\\
\hline
$Q_3$   & \quad 6.735 (6.848)	& \quad 84 (76) 	& \quad 35 (50)			\\
\hline
$Q_4$   & \quad 6.705 (6.737)	& \quad 59 (50) 	& \quad 71 (147)		\\
\hline \hline
\end{tabular}
\end{center}
\caption{Parameters for the four qubits.  Each was designed with a different target $\kappa_r$ in order to test the tradeoff between damping and measurement speed.  Measured values are in parentheses.}
\label{Table:parameters}
\end{table}

We use a multi-tone signal, generated with a custom microwave frequency arbitrary waveform generator, to simultaneously probe each of the measurement resonators \cite{Chen:readout}. Each qubit imparts a state dependent phase shift to one frequency component of the measurement pulse. The phase shifted signal is amplified by a Josephson parametric amplifier (paramp) with near quantum limited performance over a bandwidth of 600~MHz and a 1 dB compression point of approximately -107\,dBm \cite{White:IMPA}. The large bandwidth and saturation power of the amplifier was critical in our ability to simultaneously measure all four qubits. The signal is weakly filtered by a 250\,MHz Gaussian filter before it is digitized, and the amplitudes and phases for each frequency component are extracted.  For each frequency this yields a point in the quadrature (IQ) plane that depends on the state dependent phase shift imparted by the qubit.

Each measurement pulse consists of a very short (25-50\,ns) high-power transient to ring up the resonator as quickly as possible, followed by a short sustain pulse (150\,ns), as shown in Fig.\,\ref{fig:readoutpulse}(a).  The resonator rings down naturally, with a decay rate $\kappa_r = \omega_r / Q_r$, which is the slowest part of the sequence as shown in Fig.\,\ref{fig:readoutpulse}(b). The qubit can only be coherently manipulated again after several resonator decay time constants.

The IQ points for many single-shot measurement events in which the qubit was prepared in the $\ket{0}$ and $\ket{1}$ states are shown in Fig.~\ref{Fig:Figure3}. Each point is generated by integrating from the beginning of the demodulated measurement signal ($\textrm{time}=0$ in Fig.\,\ref{fig:readoutpulse}). Shots are recorded as $\ket{0}$ or $\ket{1}$ according to which cloud's centroid is nearest to the resulting IQ value.

At equilibrium, our qubits have a 5-10\% probability to be in the excited state.  To separate this effect from other sources of error, we use heralding \cite{Johnson:heralding}; we begin each sequence with an initial measurement and discard trials where the qubit does not start in the ground state.

Focusing on a single qubit-resonator pair $Q_2$, we measure the qubit and resonator frequencies spectroscopically, and then find the probe frequency for which the two IQ clouds corresponding to the qubit ground and excited states are maximally separated. All subsequent measurement pulses on this qubit use this frequency. We calibrated the photon number occupation in the resonator by measuring the AC Stark shift of the qubit\,\cite{Schuster:acStarkDephasing}.

Measuring the qubits' $T_1$ versus frequency, we find that in all four designs there was no observable suppression of $T_1$ at the smallest $\Delta/2\pi$ achievable (approximately 800\,MHz) indicating that the filter successfully isolated the qubits from the environment. All four qubits were operated with $T_1$ values between 10\,$\mu$s and 12\,$\mu$s.

\begin{figure}
\begin{centering}
\includegraphics[width=8cm]{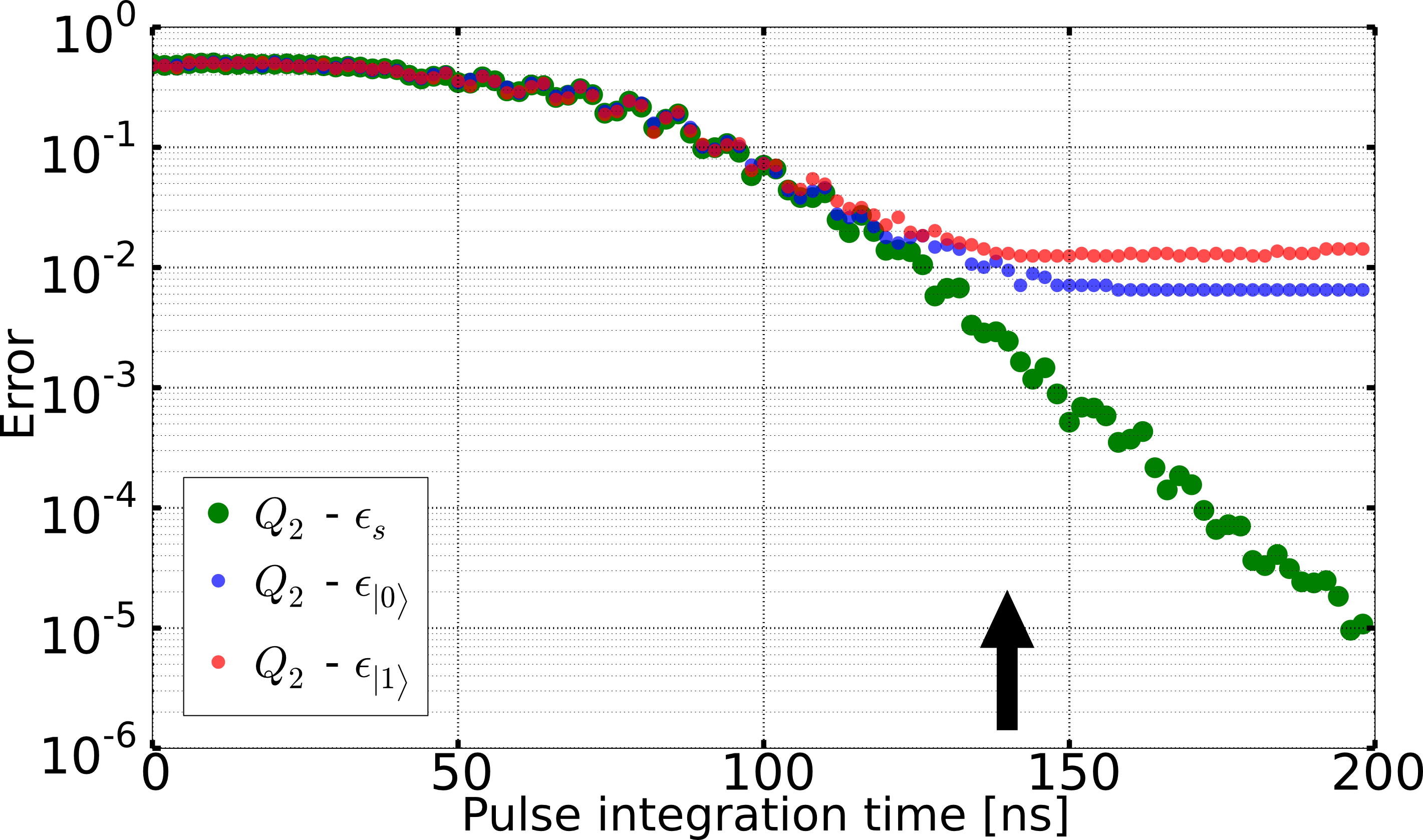}
\par\end{centering}
\caption{(color online) Measurement errors versus pulse integration time for one qubit. Large (green) circles show the separation errors \mbox{$\epsilon_s = 1 - F_s(t)$} while the dark (blue) and light (red) small circles show $\epsilon_{\ket{0}}(t)$ and $\epsilon_{\ket{1}}(t)$ respectively. The vertical arrow indicates the time slice at 140\,ns represented in Fig.\,\ref{Fig:Figure3}.}
\label{Fig:Figure4}
\end{figure}

In Fig.\,\ref{Fig:Figure3} we show results for a single qubit at a single integration time. For each point we prepare either $\ket{0}$ or $\ket{1}$ with the absence or presence of a $\pi$-pulse, and then turn on the measurement. We integrate the measurement signal for 140\,ns beginning at the start of the pulse when there are zero photons in the resonator. We characterize the measurement in two ways. First, we consider the ``separation fidelity'' $F_s$, which characterizes the distinguishability of the Gaussian fits to the IQ clouds of the two qubit states. Because of the finite separation and widths of the clouds, a point drawn from the IQ distribution for either state may be erroneously identified as the other state. We define $F_s$ as the probability that a point drawn from the fitted distribution for either state is correctly identified. Here we find $F_s=99.8\%$. Second, we define the total measurement fidelities $\{F_{\ket{x}}\}$ as the probability that a qubit prepared as $\ket{x}$ is correctly identified. This includes unwanted qubit state transitions during the non-zero duration of the measurement. While these errors arise fundamentally from the qubit, we regard them as measurement errors here because they can be reduced with faster measurement. We find $F_{\ket{0}} = 99.3\%$ and $F_{\ket{1}} = 98.7\%$.

While separation fidelity is improved by collecting more scattered photons, this requires longer measurement and thus incurs more qubit errors. To fully characterize this time dependence we measured the separation and total fidelities as functions of integration time, as shown in Fig.\,\ref{Fig:Figure4}. We use the same procedure as in Fig.\,\ref{Fig:Figure3} but vary the upper limit in the time integration to generate a time series of IQ clouds from which we extract $F_s(t)$, $F_{\ket{0}}(t)$, and $F_{\ket{1}}(t)$. We used $F_s(t)$ as an empirical optimal window and re-integrate the data weighted by this window. For clarity we plot the errors, defined as $\epsilon \equiv 1-F$, instead of the fidelities. The separation fidelity reaches $99\%$ at 124\,ns after the pulse start, and improves exponentially with increasing integration time.

The data with near constant slope shows that, after the initial transient of the measurement pulse, $\epsilon_s(t)$ decreases at a rate of approximately one decade per 25\,ns. This rate depends on the ratio between the detected photon flux and the system noise (SNR). Loss of any scattered photons before they are detected lowers the SNR. As each scattered photon carries partial information on the qubit state it also causes qubit dephasing. This provides a way to measure the fraction of lost photons: we compare the experimental SNR to the dephasing induced by the measurement \cite{_supplementary, Makhlin:quantumStateEngineering}. In this way we find a quantum efficiency of -9\,dB, of which -3\,dB can be attributed to using a phase insensitive amplifier\,\cite{_supplementary}. We note that, as it would improve only the steady state SNR but not the transient response, increasing the quantum efficiency would improve the measurement performance only slightly.

The state errors decrease along with the separation error for the first 100\,ns before they begin to saturate. This saturation can be explained by considering two deleterious qubit state transition processes. We have measured that in equilibrium our qubits experience upward $\ket{0} \rightarrow \ket{1}$ transitions with a rate of $\Gamma_{\uparrow} \approx 1/100\,\mu\textrm{s}$ which result in excited state populations of 5 to 10\%. These transitions lead to state preparation errors; with 500\,ns between the heralding and final measurements, we expect 0.5\% re-population of the excited state before the start of the final measurement. This nearly explains the saturation of $F_{\ket{0}}$ at 99.3\%. The second error process is the usual qubit energy relaxation; a qubit transition before the halfway point of the measurement leads to an error. With a measurement time of 140\,ns and $T_1 = 10\,\mu\textrm{s}$ we expect an extra 0.7\% loss in excited state population yielding an expected limit of 98.8\%. This agrees well with the measured $F_{\ket{1}}$ saturation at 98.7\%.

\begin{figure}
\begin{centering}
\includegraphics[width=8cm]{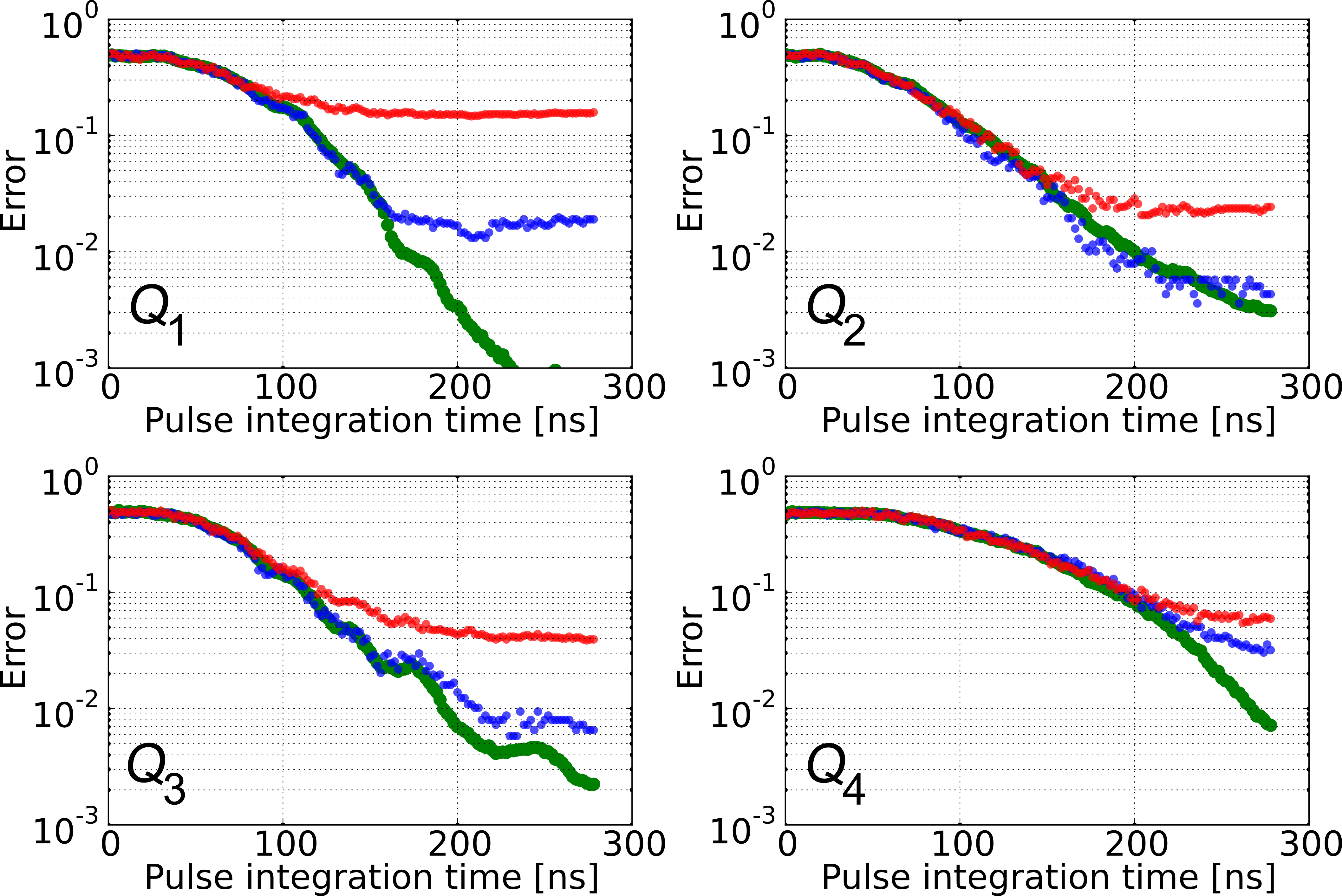}
\par\end{centering}
\caption{(color online) Simultaneous measurement of four qubits. Separation and actual state fidelities are shown as in Fig.\,\ref{Fig:Figure4}. All four qubits exhibit fast measurement, with three of them reaching 99\% fidelity in 200\,ns. Small ripples on qubits $Q_1$ and $Q_3$ were caused by spectral leakage.}
\label{Fig:Figure5}
\end{figure}

We also measured all four qubits simultaneously, as shown in Fig.~\ref{Fig:Figure5}. Three of the four qubits reached 99\% separation fidelity within 200\,ns. The fourth device, which had the most conservative $\kappa_r T_1$ product, reached 99\% separation fidelity at 266\,ns. In order to prevent saturation of the paramp with four simultaneous measurement tones, we reduced the drive power relative to the single qubit measurement. This required an increase in the measurement time which led to slightly lower fidelity than was achieved with a single qubit.

For qubits $Q_2$ and $Q_4$ the performance is nearly as good as for the single qubit case. The small degradation of performance comes from increased qubit transitions during the longer measurement time. Qubits $Q_1$ and $Q_3$ show lower $F_{\ket{1}}$. As shown in the inset of Fig.\,\ref{Fig:Figure1} the measurement resonators for qubits $Q_1$ and $Q_3$ are closely spaced in frequency (13\,MHz). This close spacing adversely affects the frequency discrimination step of the measurement via spectral leakage, leading to increased measurement error. In addition the measurement photons induce large qubit frequency shifts (200-300\,MHz) via the AC Stark effect. This causes the qubits to cross through resonance with material defects and lose $\ket{1}$ population. We were able to work around this problem with careful choice of operating frequency in qubits $Q_2$, $Q_3$, and $Q_4$, but limited total available frequency space led to degraded performance in $Q_1$. This problem would be substantially mitigated in devices constructed with epitaxial Al films grown on plasma cleaned substrates \cite{Megrant:highQ} as this was shown to produce qubit frequency spectra with a significant reduction in defects \cite{Barends:XMon}.

In conclusion, we have demonstrated fast and accurate multi-qubit state measurement in superconducting qubits. Amplifier saturation power is a key metric for system performance, and further improvements in amplifiers would allow the bandpass filter design to scale to even higher numbers of qubits. This system is suitable for more complex experiments with larger numbers of qubits, and meets the threshold requirements for surface code error correction.

This work was supported by the Office of the Director of National Intelligence (ODNI), Intelligence Advanced Research Projects Activity (IARPA), through the Army Research Office grants W911NF-09-1-0375 and W911NF-10-1-0334.  All statements of fact, opinion, or conclusions contained herein are those of the authors and should not be construed as representing the official views or policies of IARPA, the ODNI, or the U.S. Government.
\bibliographystyle{apsrev}
\bibliography{references}

\end{document}


\title{Supplementary information for Fast Scalable State Measurement with Superconducting Qubits}

\author{Evan Jeffrey} \thanks{These authors contributed equally to this work} \affiliation{Department of Physics, University of California, Santa Barbara, California 93106-9530, USA}
\author{Daniel Sank} \thanks{These authors contributed equally to this work} \affiliation{Department of Physics, University of California, Santa Barbara, California 93106-9530, USA}
\author{J.Y. Mutus} \affiliation{Department of Physics, University of California, Santa Barbara, California 93106-9530, USA}
\author{T.C. White} \affiliation{Department of Physics, University of California, Santa Barbara, California 93106-9530, USA}
\author{J. Kelly} \affiliation{Department of Physics, University of California, Santa Barbara, California 93106-9530, USA}
\author{R. Barends} \affiliation{Department of Physics, University of California, Santa Barbara, California 93106-9530, USA}
\author{Y. Chen} \affiliation{Department of Physics, University of California, Santa Barbara, California 93106-9530, USA}
\author{Z. Chen} \affiliation{Department of Physics, University of California, Santa Barbara, California 93106-9530, USA}
\author{B. Chiaro} \affiliation{Department of Physics, University of California, Santa Barbara, California 93106-9530, USA}
\author{A. Dunsworth} \affiliation{Department of Physics, University of California, Santa Barbara, California 93106-9530, USA}
\author{A. Megrant} \affiliation{Department of Physics, University of California, Santa Barbara, California 93106-9530, USA}
\author{P.J.J. O'Malley} \affiliation{Department of Physics, University of California, Santa Barbara, California 93106-9530, USA}
\author{C. Neill} \affiliation{Department of Physics, University of California, Santa Barbara, California 93106-9530, USA}
\author{P. Roushan} \affiliation{Department of Physics, University of California, Santa Barbara, California 93106-9530, USA}
\author{A. Vainsencher} \affiliation{Department of Physics, University of California, Santa Barbara, California 93106-9530, USA}
\author{J. Wenner} \affiliation{Department of Physics, University of California, Santa Barbara, California 93106-9530, USA}
\author{A. N. Cleland} \affiliation{Department of Physics, University of California, Santa Barbara, California 93106-9530, USA}
\author{John M. Martinis} \affiliation{Department of Physics, University of California, Santa Barbara, California 93106-9530, USA}

\begin{abstract}
Here we present theoretical and technical details on the design of the bandpass filter and measurement resonators used in the experiment. We also describe the method used to measure the quantum efficiency of the detector. Finally we show a full diagram of the experimental set-up.
\end{abstract}

\maketitle

\section{Environmental Limit of Qubit Coherence}

In this section we present analytic and numerical models of the response-lifetime product $\kappa_r T_1$. Because the qubit is nearly harmonic we can use linear circuit theory to calculate $T_1$ of the excited state $\ket{1}$ \cite{Esteve:dissipation}. We calculate the $Q$ of an equivalent linear circuit element and then write it in terms of $T_1$ using $Q_q \equiv \omega_q T_1$.

\subsection{Theory}

\begin{figure}
\begin{centering}
\includegraphics[width=8cm]{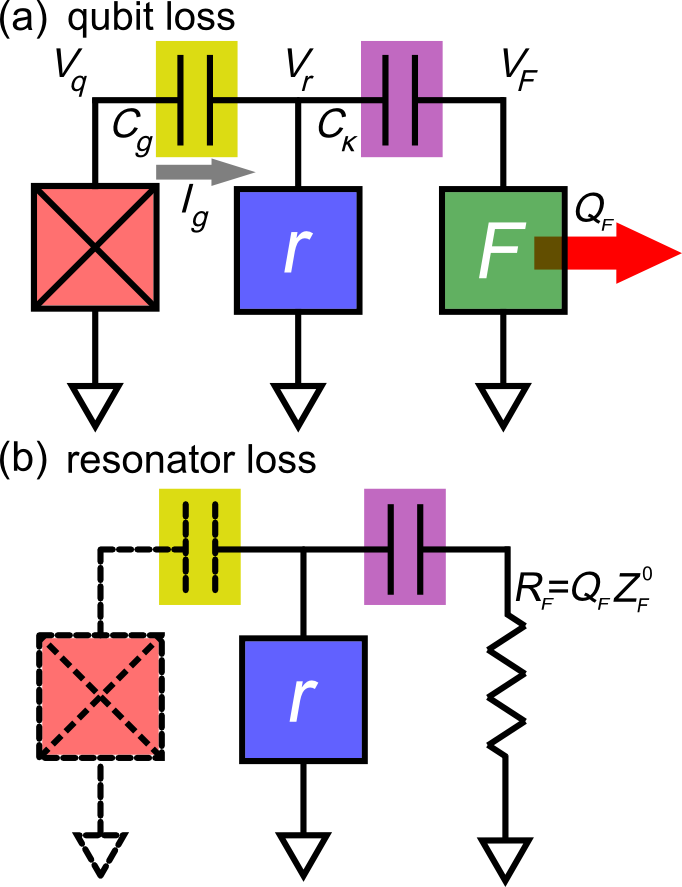}
\par\end{centering}
\caption{(color online) Lumped element circuit model of the measurement system. The qubit (cross), measurement resonator ($r$), and filter ($F$) are connected through coupling capacitors. (a) For qubit loss we assume the filter is the only lossy element, so system energy only leaves through the finite $Q_F$ of the filter. (b) For calculation of $Q_r$ we work at the measurement resonator frequency. The measurement resonator and filter are on resonance, so the filter impedance is nearly real and modelled as a resistor. The qubit and coupling capacitor $C_g$, indicated with dotted lines, are lossless and therefore ignored.}
\label{Fig:Figure1}
\end{figure}

We first present an analytic calculation. A diagram of the theoretical model is shown in Fig.\,\ref{Fig:Figure1}. The quality factor of the qubit $Q_q$ is defined as \begin{equation}
Q_q \equiv \frac{\textrm{energy stored in qubit}}{\textrm{energy lost per radian}} \, . \label{eq:definitionOfQ_q} \end{equation}
The energy stored in the qubit is $E_q =\frac{1}{2} C_q \left| V_q \right| ^2$ where $C_q$ is the qubit capacitance and $V_q$ is the voltage amplitude at the qubit node as indicated in Fig.\,\ref{Fig:Figure1}. Assuming that the only lossy element in the system is the filter, we use the definition of the filter quality factor $Q_F$ to write \begin{equation}
\textrm{energy lost per radian} = E_F / Q_F \label{eq:E_l} \, , \end{equation}
where $E_F = \frac{1}{2}C_F\left| V_F \right|^2$ is the energy stored in the filter. Inserting Eq.\,(\ref{eq:E_l}) into (\ref{eq:definitionOfQ_q}) yields \begin{equation}
Q_q = Q_F \frac{C_q}{C_F} \left| \frac{V_q}{V_F} \right|^2 \, , \label{eq:Q_q} \end{equation}
where $C_F$ and $V_F$ are the filter capacitance and voltage amplitude. See Fig.\,\ref{Fig:Figure1}(a). To compute the ratio $V_q/V_F$ we use voltage division. We make the crucial observation that to calculate the qubit damping we must analyze the circuit impedances \emph{at the qubit frequency}. Because the qubit is off resonance from the measurement resonator, the measurement resonator's impedance is low and we assume \mbox{$Z_g \gg Z_{r}$}. Therefore with voltage $V_q$ across the qubit, we have a current \mbox{$I_g=V_q/Z_g$} flowing through $C_g$. By similar reasoning \mbox{$Z_{\kappa} \gg Z_r$}, so most of the current $I_g$ flows through the measurement resonator. This gives $V_r = I_g Z_r = V_q Z_r / Z_g$. Using similar arguments to work through each stage of the circuit we arrive at \begin{equation}
\frac{V_q}{V_F} = \frac{Z_g Z_{\kappa}}{Z_r Z_F} \, . \label{eq:voltageRatio} \end{equation}
Note the shunt impedances in the denominator and the coupling impedances in the numerator.

Next we compute $Z_r$ and $Z_F$ in terms of their characteristic impedances. The impedance of a parallel harmonic mode is \begin{equation}
\frac{1}{Z} = \frac{i}{Z^0} \frac{2\delta x + \delta x^2}{1+\delta x} \, , \label{eq:resonanceImpedance} \end{equation}
where $\delta x \equiv (\omega - \omega_0)/\omega_0$, $\omega_0$ is the resonance frequency, and $Z^0$ is the characteristic impedance of the resonance (equal to $\sqrt{L/C}$ for a parallel LC). Inserting Eq.\,(\ref{eq:resonanceImpedance}) into (\ref{eq:voltageRatio}) we get \begin{equation}
\left| \frac{V_q}{V_F} \right| = \frac{\left| Z_g \right| \left| Z_{\kappa} \right|}{Z_r^0 Z_F^0} \left( \frac{2\delta x + \delta x^2}{1 + \delta x} \right) ^2 \, , \label{eq:voltageRatio_1} \end{equation}
where here $\delta x \equiv (\omega_q - \omega_r) / \omega_r$, $\omega_r$ is the measurement resonator frequency, and we assume the measurement resonator and filter have the same resonance frequencies. Inserting Eq.\,(\ref{eq:voltageRatio_1}) into (\ref{eq:Q_q}) yields \begin{equation}
Q_q = Q_F \frac{C_q}{C_F}\left( \frac{\left| Z_g \right| \left| Z_{\kappa} \right|}{Z_r^0 Z_F^0} \right)^2 \left( \frac{2\delta x + \delta x ^2}{1 + \delta x} \right)^4 \, . \label{eq:Q_q_circuitOnly} \end{equation}

Equation (\ref{eq:Q_q_circuitOnly}) expresses $Q_q$ in terms of circuit element values, but to produce a more useful design formula we must eliminate $Z_{\kappa}$ in favor of $Q_r$. To calculate $Q_r$ we work at the measurement resonator frequency. With the measurement resonator and filter assumed to be nearly on resonance the filter appears as a pure resistance \mbox{$R_F=Q_F Z_F^0$}, as shown in Fig.\,\ref{Fig:Figure1}(b). We assume the qubit to be lossless so the filter resistance sets $Q_r$. Using a method similar to that which led to Eq\,(\ref{eq:Q_q_circuitOnly}) we find \begin{equation}
Q_r = \frac{\left|Z_{\kappa} \right|^2}{R_F Z_r^0} = \frac{\left| Z_{\kappa} \right|^2}{Q_F Z_F^0 Z_r^0} \, . \label{eq:Q_r} \end{equation}
Substituting Eq.\,(\ref{eq:Q_r}) into (\ref{eq:Q_q_circuitOnly}) and using \mbox{$Q_q = \omega_q T_1$} and \mbox{$Q_r = \omega_r / \kappa_r$} we find \begin{equation}
\kappa_r T_1 = Q_F^2 \left( \frac{\omega_r}{\omega_q} \right)^2 \left( \frac{C_q}{C_g} \right) ^2 \frac{Z_q^0}{Z_r^0} \left( \frac{2\delta x + \delta x^2}{1 + \delta x} \right)^4 \, . \label{eq:Q_q_lumped} \end{equation}
Equation (\ref{eq:Q_q_lumped}) is most useful when comparing with results from numerical circuit simulators and when choosing values for the actual circuit hardware. For the present experiment in which we use $\lambda/4$ resonators it is convenient to use the relation between the filter characteristic impedance and the line impedance $Z_r^0 = (4/\pi)Z_0$ resulting in \begin{equation}
\kappa_r T_1 = \frac{\pi}{4} Q_F^2 \left( \frac{\omega_r}{\omega_q} \right) ^2 \left( \frac{C_q}{C_g} \right)^2 \frac{Z_q^0}{Z_0} \left( \frac{2 \delta x + \delta x^2}{1 + \delta x} \right)^4 \, . \label{eq:Q_q_final} \end{equation}
We used Eq.\,(\ref{eq:Q_q_final}) as our design formula.

For an equation applicable to other physical systems we eliminate capacitances and impedances in favor of coupling constants. Using the standard formula for capacitive coupling between harmonic modes \begin{equation}
g = \frac{1}{2} \frac{C_g}{\sqrt{C_q C_r}} \sqrt{\omega_q \omega_r} \, , \end{equation}
and keeping only the leading order in $\delta x$ we can re-express Eq.\,(\ref{eq:Q_q_final}) as \begin{equation}
\kappa_r T_1 = 4 \frac{\Delta^4}{g^2 \omega_q^2 / Q_F^2} = \left( \frac{\Delta}{g}\right)^2 \left(\frac{\omega_r}{\omega_q} \frac{2 \Delta}{\omega_r / Q_F} \right)^2 \, , \label{eq:T1} \end{equation}
where $\Delta \equiv \omega_q - \omega_r$, and $\omega_r/Q_F$ is the filter bandwidth. Equation (\ref{eq:T1}) provides the link between measurement time and qubit coherence. With our design parameters $Q_F=30$, \mbox{$\Delta/2\pi = 800\,\textrm{MHz}$}, \mbox{$g/2\pi=90\,\textrm{MHZ}$}, \mbox{$\omega_r/2\pi=6.8\,\textrm{GHz}$}, and \mbox{$\omega_q/2\pi=6\,\textrm{GHz}$} we get \mbox{$\kappa_r T_1 = 5050$.} An engineered leakage rate of \mbox{$\kappa_r = 1/50\,\textrm{ns}$} gives $T_1 = 250\,\textrm{us}$. We designed our four $\kappa_r$ values to range from 1/12\,ns to 1/71\,ns.

\begin{figure}
\begin{centering}
\includegraphics[width=8cm]{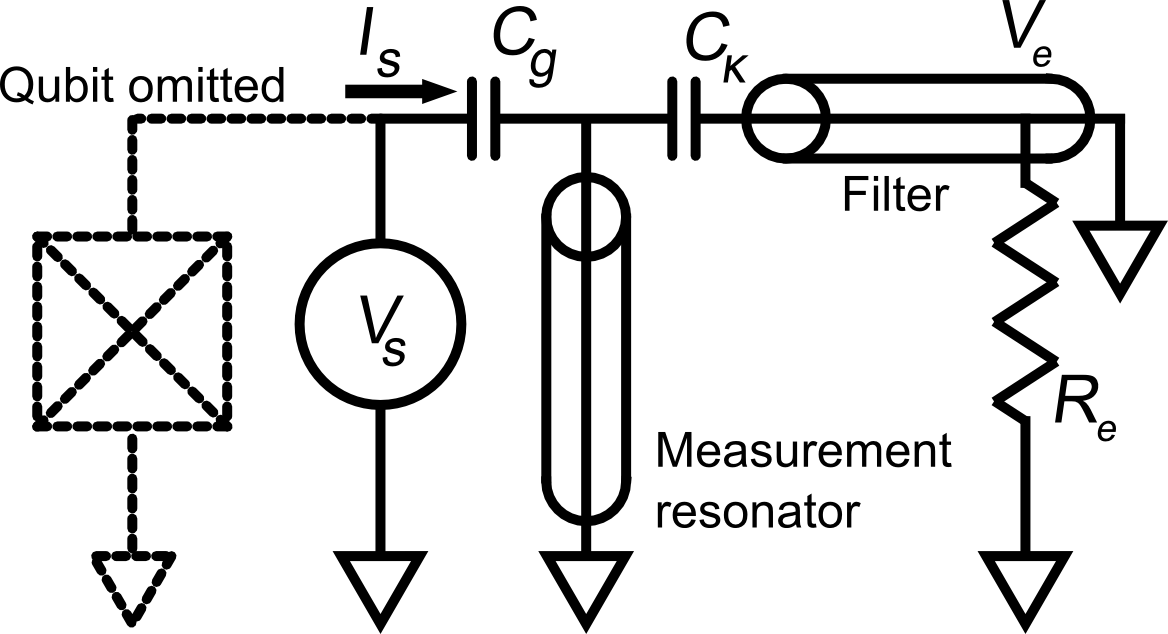}
\par\end{centering}
\caption{Circuit model used in SPICE simulation. Elements shown in solid line were simulated. The resistor $R_e$ models the 50$\,\Omega$ environment imposed by the amplification chain. The impedance of the circuit shown in solid line is measured by sourcing $V_s$ and measuring $I_s$.}
\label{Fig:Figure2}
\end{figure}

\subsection{Numerics}

\begin{figure}
\begin{centering}
\includegraphics[width=8cm]{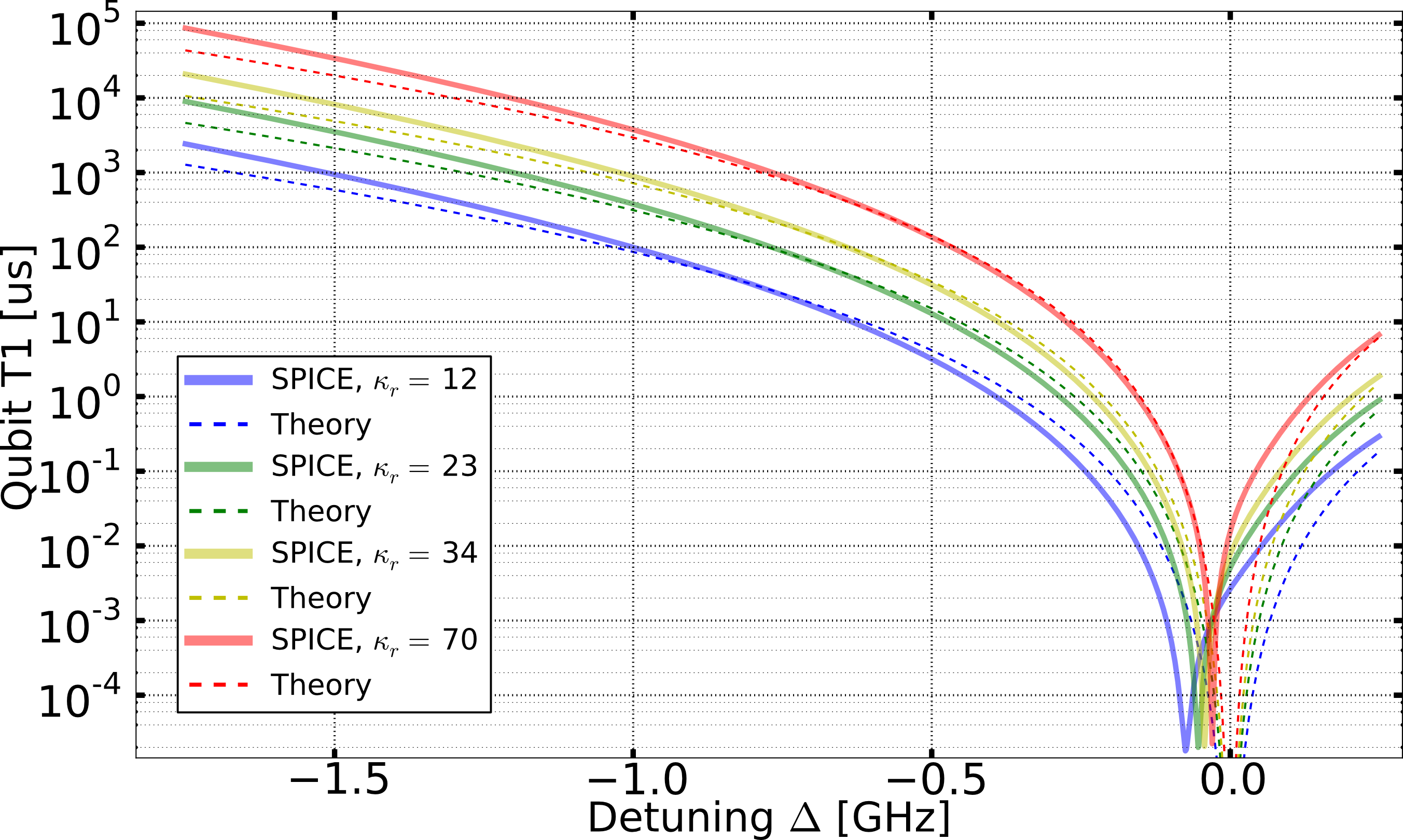}
\par\end{centering}
\caption{(color online) Analytic (Eq.\,\ref{eq:Q_q_circuitOnly}) and numerical (SPICE) qubit $T_1$ limits for several values of $\kappa_r$.}
\label{Fig:Figure3}
\end{figure}

We compared Eq.\,(\ref{eq:Q_q_final}) against a numerical simulation of the circuit in SPICE \footnote{www.linear.com/designtools/software}. The circuit model is shown in Fig.\,\ref{Fig:Figure2}. The quality factor of the qubit is determined in a simple two step procedure. First, we replace the qubit with a voltage source. We activate the voltage source with an amplitude $V_s$ at frequency $\omega$ and record the complex current $I_s$ flowing into the rest of the circuit. The admittance of the circuit external to the qubit is then \begin{equation}
Y_e(\omega) = I_s / V_s \, . \end{equation}
Second, we compute the $T_1$ of the qubit as \cite{Esteve:dissipation} \begin{equation}
T_1 = C_q / \left| \mathrm{Re} Y_e(\omega_q) \right| \, . \end{equation}
Results of the simulation with corresponding predictions from Eq.\,(\ref{eq:Q_q_final}) are shown in Fig.\,\ref{Fig:Figure3}. We plot the $T_1$ limit versus detuning between the qubit and measurement resonator for several values of $Q_r$. We note that the simple linear theory agrees very well with the numerical result up to $\Delta/(2\pi) \approx 1 \, \textrm{GHz}$. The disparity at larger detunings probably comes from the assumption, made in deriving Eq.\,(\ref{eq:voltageRatio}), that the coupling capacitance impedances $Z_g$ and $Z_{\kappa}$ are much larger than the resonator impedances at the qubit frequency.

\section{Quantum Efficiency}

The rate of separation fidelity improvement during the equilibrium part of the measurement increases with increasing flux of detected measurement photons. Each measurement photon carries information about the qubit state and therefore incurs dephasing of the qubit \cite{Makhlin:quantumStateEngineering}. This results in a direct relationship between the separation of the measured IQ clouds and the qubit phase coherence (ignoring any additional decoherence channels in the qubit)\begin{equation}
\left| \rho_{10} \right| = \exp \left[ -\frac{s^2}{8 \sigma^2} \right] \, . \label{eq:measurementDephasing} \end{equation}
Here $\rho_{10}$ is the amplitude of the off-diagonal elements of the qubit density matrix, $s$ is the distance between the centers of the $\ket{0}$ and $\ket{1}$ IQ clouds, and $\sigma$ is their widths (assumed to be equal). Equation (\ref{eq:measurementDephasing}) provides a means of determining the fraction of photons lost to dissipation in the measurement system. Lost photons decohere the qubit, but do not contribute to the separation of the IQ clouds. Therefore, by measuring the cloud separation and the dephasing induced on the qubit, we can extract the fraction of photons lost in the measurement process. We found a photon collection efficiency of -9\,dB, or 12.6\%. We attribute -3\,dB to using a parametric amplifier (paramp) in phase preserving mode \cite{Caves:amplifiers}, -2\,dB from infrared filters used on the signal output line, and the rest to a combination of losses in microwave switches, circulators, and connectors. There is also a small amount of added noise from the HEMT amplifier due to the finite gain of the paramp.

\section{Experimental Set-up}

\begin{figure*}
\includegraphics[width=\textwidth]{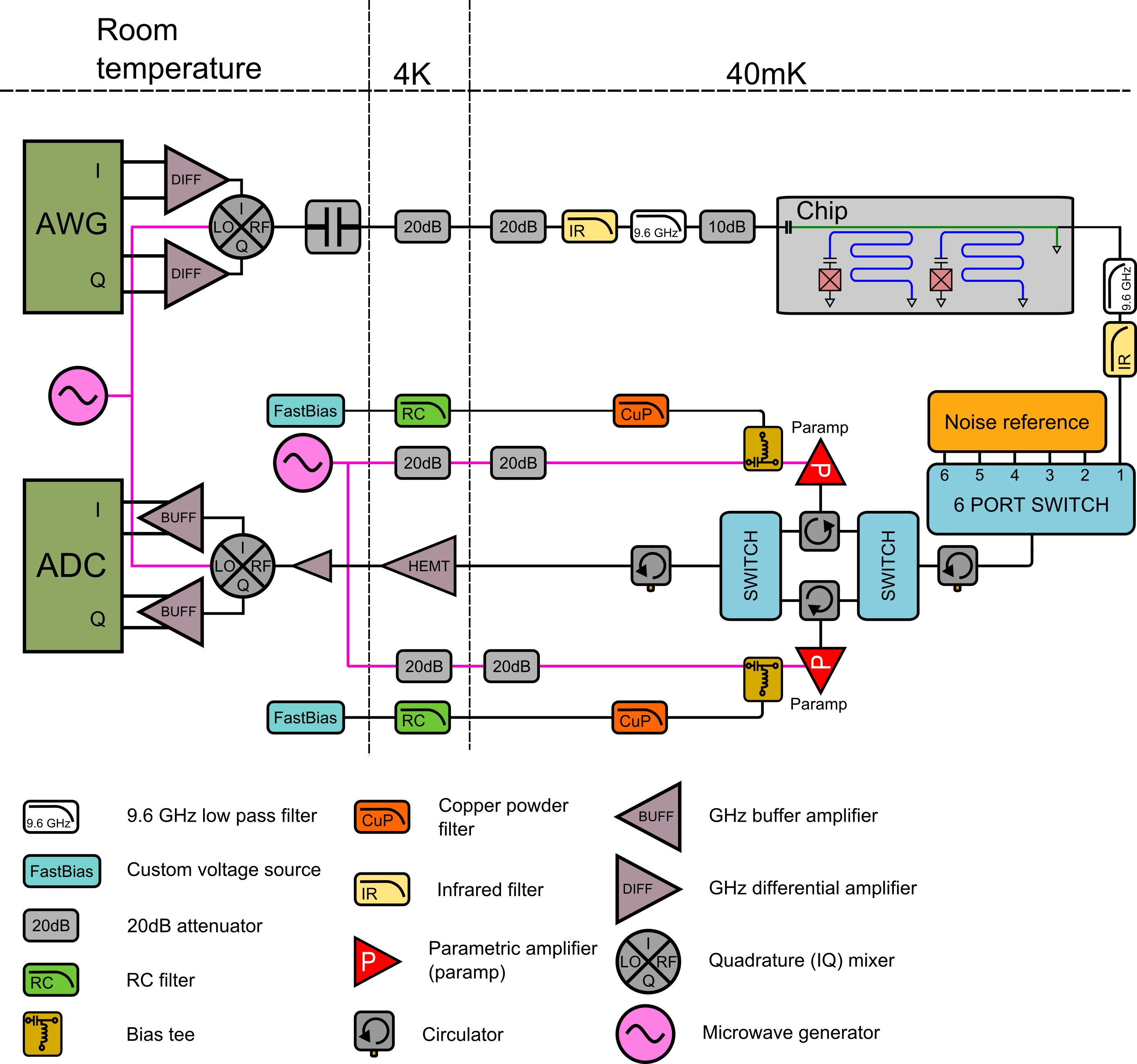}
\caption{The experimental set-up. Only components used for state measurement are shown. Pulses are generated by the AWG and mixed to gigahertz frequencies. Cold attenuators, microwave filters, and infra-red filters prevent noise and thermal radiation from reaching the qubits. The transmitted signal is directed through switches to one of two paramps. This allows switching between multiple samples, noise references, and paramps. The signal is further amplified by the HEMT and room temperature amplifiers and digitized. The paramp flux bias is generated by a custom voltage source and filtered by RC and copper powder filters.}
\label{Fig:wiringDiagram}
\end{figure*}

Here we describe the experimental set-up. A schematic is shown in Fig.\,\ref{Fig:wiringDiagram}. Measurement pulses are generated through sideband mixing. A custom dual channel 14-bit 1\,GS/s arbitrary waveform generator (AWG) generates 20-200\,MHz signals which are mixed with a local oscillator (LO) to generate shaped pulses at GHz frequencies. The AWG signal is a superposition of frequencies, one for each measurement resonator, so that the signal sent to the chip consists of four frequency multiplexed measurement pulses. The signal arriving at the chip is mostly reflected by the input capacitor of the bandpass filter, and only a small fraction enters the filter. Each one of the frequency multiplexed pulses is then phase shifted by one of the measurement resonators before leaving the chip through the output port. The small input capacitor ensures only a small fraction of the phase shifted signal is lost by exiting the chip through the input port. After leaving the chip the signal passes through a series of filters, switches and isolators before it is amplified by a parametric amplifier. The signal is then further amplified by a high mobility electron transistor (HEMT) amplifier and room temperature amplifiers before it is down-mixed to MHz frequencies, digitized and recorded by a custom analog to digital converter (ADC). Digital processing then separates the signal into its frequency components and extracts the phase shifts for each component.

\bibliographystyle{apsrev}
\bibliography{references}